# Effect of synthesis conditions on properties of barium titanate piezoelectric ceramics

**Mojtaba Biglar[1], Magdalena Gromada[2], Feliks Stachowicz[3], Tomasz Trzepieciński[4]**

[1, 3, 4]*Rzeszow University of Technology, Rzeszów, Poland*
[2]*Institute of Power Engineering, Ceramic Department CEREL, Research Institute, Boguchwała, Poland*

Email: *m_biglar@prz.edu.pl, gromada@cerel.pl, stafel@prz.edu.pl, tomtrz@prz.edu.pl*

**Abstract:**

The powder of barium titanate ($BaTiO_3$) synthesized by solid state was characterised by spherical grains of the average size about 500 nm. The double mechanical activation and calcination of raw materials was necessary to receive the monophase material. In order to improve the mouldability, the obtained barium titanate powder was granulated in spray drier. Pellets and beams from $BaTiO_3$ granulate were manufactured by both uniaxial and isostatic pressing. The properties of barium titanate material at each stage of its fabrication (powder, granulate, sintered material) influencing on its application for the stacked-disk multilayer actuator were determined. Particularly, the four parameters of $BaTiO_3$ sinter affecting on the usability properties of actuators, not found before in the literature, were estimated.

INTRODUCTION

Barium titanate, revealing both ferroelectric and piezoelectric properties, is a very attractive material in the field of electroceramics and microelectronics (Vijatović et al., 2008). $BaTiO_3$ is an excellent choice for many applications, such as multilayer capacitors (MLCs), multilayer actuators (MLAs) and energy storage devices (Miot et al., 1995; Kao and Yang, 1999). These applications require materials with a good density, a high dielectric constant and a low loss factor. Electric properties of ceramic material used to be very sensitive to its microstructure. Synthesis of materials with these properties needs highly pure, fine grained ceramics, which may be obtained from homogeneous reproducible powders of spherical particles (Miot et al., 1995; Kholodkova et al., 2012).

In the literature, one can found many types of actuators: piezoelectric ceramics, magnetostrictors and shape memory alloys but only piezoelectric actuators offer the combination of large generative forces, sub-millisecond response time, high volumetric efficiency and low cost while utilizing relatively simple E-field control (Yoshikawa and Shrout, 1993). Choi et al. (2008) underline that multilayer ceramic actuators have been extensively explored because their assets are a rapid operation, a low power consumption, a high precision control and little noise, compared with conventional electromagnetic actuators. Two types of multilayer actuators (MLAs) are considered, namely, stacked-disk and co-fired (Yoshikawa and Shrout, 1993). Actuators made from fine grain ceramic are expected to have improved reliability, higher driving fields, and lower driving voltages (from thinner layers in stacked or co-fired actuators) over devices fabricated from conventional materials (Hackenberger et al., 1998).

The properties of barium titanate have been the subject of study of many authors. It is well known that the features of $BaTiO_3$ powders and sinters strongly depend on the synthesis procedure and sintering conditions (Vijatović et al., 2008).

A lot of synthetic methods have been developed for the preparation of barium titanate powders. Both the conventional solid state reaction methods and the chemical methods such as sol gel, co-precipitation, hydrothermal and mechanochemical used to prepare barium titanate have been described in the literature. Large scale production is based on solid state reactions of mixed powders $BaCO_3$ and $TiO_2$ at high temperatures. However, this procedure results in agglomeration, poor chemical homogeneity and formation of secondary phases which harms the electrical properties of $BaTiO_3$ (Othman et al., 2014; Kim et al., 2009). However, the solid state method has the advantages of low production cost, precise stoichiometry control, operating simplicity and possibility of its application on mass production (Kao and Yang, 1999; Kholodkova et al., 2012).

In a connection of solid state method disadvantages, searching for new suitable routes for commercial $BaTiO_3$ synthesis was conducted in literature. In these methods wet-chemical techniques at low temperature, less than 100°C, are used for obtaining of 50-60 nm particles in small number of stages (Kholodkova et al., 2012). Nguyen et al. (2007) have developed another soft chemistry route for the synthesis of barium titanate: the hydrothermal synthesis, which according to authors is one of the best methods for producing metal oxide particles. It has many advantages such as highly crystalline particles, low-cost starting materials, low temperature treatment and simple procedure. Consequently, it allows a better control of the powder properties.

The main aim of this paper is to find out if the barium titanate powder fabricated by solid state may be successful utilised for manufacturing the general part of the stacked-disk multilayer actuators. To this end, all properties influencing on application of obtained material for the multilayer actuator were determined and discussed.

**EXPERIMENTAL**

The powder of barium titanate was manufactured from the $TiO_2$ (99% of purity, Kronos) and $BaCO_3$ (99,5% of purity, Chempur). The raw materials were mixed with isopropyl alcohol and milled in the mixer mill for two hours. The obtained powder was calcined in the electrical furnace for 8 hours at 1100°C. The double milling and calcination was necessary to obtain monophase material. The X-ray powder diffraction patterns were recorded by X'PertPROX diffractometer equipped with PIXcel detector (Cu Kα radiation). The microstructure of powder was evaluated by means of scanning electron microscopy SEM/HITACHI S-3400N/2007. The particle size distribution of $BaTiO_3$ powder was determined using Mastersize 2000 (Melvern). The specific surface area and the pore size of fabricated powder were determined by the low-temperature nitrogen adsorption method (77 K, Quantachrome Autosorb-1).

In order to improve the mouldability, the obtained barium titanate powder was granulated. The powder was mixed with deionized water in the ball mill for 30 minutes and then dispex, oil emulsion and polyvinyl alcohol were added. The granulation process was performed in spray drier (Niro) at the inlet temperature of 220°C, the outlet temperature of 80°C and the spray pressure was equal to 40 mm water column. The basic properties of received granulate: the SEM microstructure, the phase composition, the specific surface area and the pore size were determined.

Pellets from $BaTiO_3$ granulate were manufactured by both uniaxial and isostatic pressing. Then the green pellets were sintered in the electrical furnace according to the five curves, determined on the base of result received from the dilatometric curve, and ensuring high density of sintered discs. The highest temperatures of sintering were 1250, 1300, 1350, 1400 and 1450°C, the dwell time - 4 hours and the rate of heating and cooling was equal to 100°C/h. Sintered pellets were denoted, respectively: BT_0, BT_1, BT_2, BT_3 and BT_4.

The parameters characterizing the extent of material sintering: the apparent density, the apparent porosity, the water absorbability were estimated for sintered pellets by method taking advantages of Archimedes law. In the next stage, pellet of the average density, sintered at 1300°C (BT_1) was considered and characterised in the direction of XRD and SEM.

Beams produced according to the same procedure as BT_1 pellets, were used for determining the coefficient of thermal expansion, the bending strength, the hardness and the fracture toughness. The coefficient of linear expansion in the range of temperature from 20 to 600°C was determined using high temperature dilatometer from BÄHR-Gerätebau GmbH company. The bending strength of BT_1 material was determined by three-point method on the sintered not polished samples. The strength testing machine INSTRON from Zwick company was applied. The hardness and fracture toughness $K_{IC}$ of sintered BT_1 was performed on Vickers hardness testing machine 430/450SVD (Wilson Hardness).

**RESULTS AND DISCUSSION**

Fig. 1 presents the XRD pattern of barium titanate powder, which fully overlaps with the standard spectrum. The spacing and width of standard and synthesized powder are very well convergent. There is the only slightly difference in the level of peaks intensity. Therefore, one can draw a conclusion that after the second mechanical activation and calcination of barium titanate, the monophase powder was obtained.

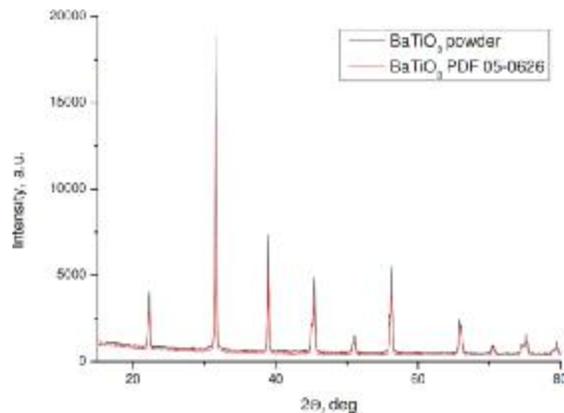

**Fig. 1. XRD pattern of BaTiO$_3$ powder together with standard spectrum**

Photo of the SEM microstructure of BaTiO$_3$ powder on the left hand side of Fig. 2 points that the received powder possesses homogeneous structure and dimension. The greater magnification of powder presented on Fig. 2b allows to estimate the grain size of barium titanate powder as 500 nm. The only a few grains visible in this photo have the mean size about 1 µm.

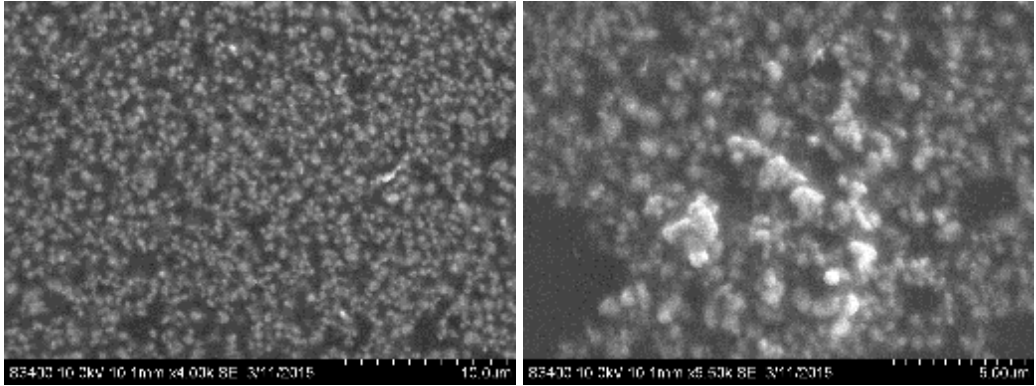
**Fig. 2. SEM microstructure of BaTiO$_3$ powder in two different magnifications**

The particle size distribution of barium titanate powder visible on Fig. 3 is bimodal and ranges from 0.3 to 80 µm. The mean particle size of this powder is equal to 1.59 µm, while the $d_{10}$ and $d_{90}$ are equal to 0.66 and 20.64 µm, respectively. The discrepancy between the result of SEM microstructure presented on Fig. 2 and the particle size distribution of barium titanate powder (Fig. 3) follows from the easiness creation of agglomerates. Such cluster of agglomerates forms particles which were measured on Mastersize 2000 apparatus and the value of the mean particle size was over three times higher than the separate grain. It should be underlined that Yoon and Lee (2004) in their research have used the commercial powder of barium titanate made by solid state of very similar, to obtained in this approach, the mean particle size 1.38 µm and a specific surface area of 2.84 m$^2$/g.

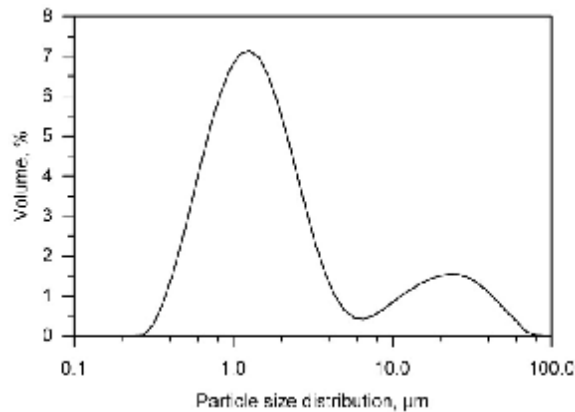
**Fig. 3. The particle size distribution of BaTiO$_3$ powder**

The specific surface area of barium titanate powder was equal to 1.87 m$^2$/g, which was a very small value. Moreover, the measurements of low-temperature nitrogen sorption showed that this material exhibited only trace porosity, which confirms insignificant value of pores volumes reaching 0.0001 cm$^3$/g.

In order to check if the granulation process did not influence on the phase composition of barium titanate, the measurement of XRD spectrum was performed. Fig. 4 presents XRD pattern of granulate, which is still very well fitted with the standard spectrum.

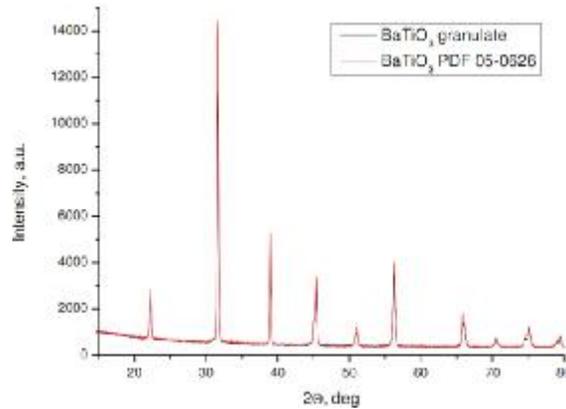
**Fig. 4. XRD pattern of BaTiO$_3$ granulate together with standard spectrum**

During the granulation process, the water slurry was sprayed in the nozzle and as a result, the filled balls of the dimension from 25 to 35 μm were created and presented on Fig. 5. The granules are built from closely compacted spherical grains of powder. The size of powder grains did not change because before the granulation process, the powder was only mixed with additives.

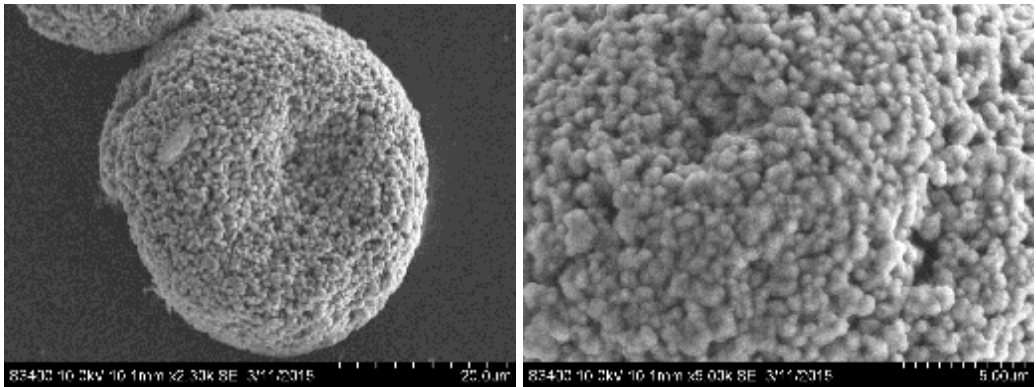
**Fig. 5. SEM microstructure of BaTiO$_3$ granulate in two different magnifications**

The specific surface area of barium titanate granulate is even smaller than in the case of powder and it is equal to 1.18 m$^2$/g. However, the barium titanate granulate possesses the pores volumes three times higher (0.0003 cm$^3$/g) than for powder, which is still very small value. The result of pores volume of granulate is in a good agreement with outcome of the SEM photos (Fig. 5), on which small amount of slightly pores are visible.

From data placed in tab. 1 follows that the values of water absorbability and the apparent porosity of sintered pellets increase together with the maximal sintering temperature. However, the values of apparent density for all pellets are quite similar and do not have any tendency. The theoretical density of sintered barium titanate calculated on the base of value of lattice constant is 6.02 g/cm$^3$ (Miot et al., 1995; Duran et al., 2002). Therefore, the ratio of the apparent density to the theoretical one (the relative density) was also estimated and placed in tab. 1. The obtained values of relative density are very high and comparable to these presented by other authors (Miot et al., 1995; Duran et al., 2002).

**Tab. 1. The apparent density, the water absorbability and the apparent porosity of sintered pellets**

| Material | Water | Apparent | Apparent | Relative |
|---|---|---|---|---|

|      | absorbability, % | density, g/cm$^3$ | porosity, % | density, % |
|------|------|------|------|------|
| BT_0 | 0.03 | 5.84 | 0.19 | 97.01 |
| BT_1 | 0.05 | 5.82 | 0.28 | 96.68 |
| BT_2 | 0.06 | 5.83 | 0.37 | 96.84 |
| BT_3 | 0.07 | 5.80 | 0.38 | 96.35 |
| BT_4 | 0.09 | 5.87 | 0.50 | 97.51 |

As it can be visible from Fig. 6, XRD pattern of BT_1 pellet is not so perfect as in the case of powder and granulate. Obviously, the material of pellet is still monophase, peaks are in a good agreement with standard, the spacing and width of standard and considered pellet are very well convergent, but there is the difference in level of peaks intensity.

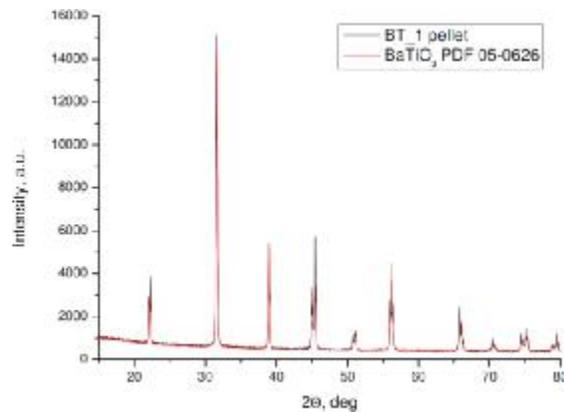

**Fig. 6. XRD pattern of BT_1 pellet together with standard spectrum**

The next parameter influencing on possibility of piezoelectric material application on actuators, is the grain size of sintered materials. To determine this feature, photos of SEM microstructure of barium titanate after sintering were prepared. Fig. 7a presents BT_1 pellet surface, where the compact densification can be observed. The microstructure is composed of both great grains of the dimension from 50 to 100 µm and smaller one of the dimension from 1 to 3 µm. The grain boundaries are very well foreshadowed and the only slender pores can be noticed in these photos, which is in agreement with result of BT_1 pellets densification presented in tab. 1. Photo of fracture surface of BT_1 pellets visible on Fig. 7b, confirms information of a very good material densification. In this dense microstructure, the small pores can be somewhere indicated of the dimension to a few micrometres. The very similar microstructure can be found in paper written by Yoon and Lee (2004).

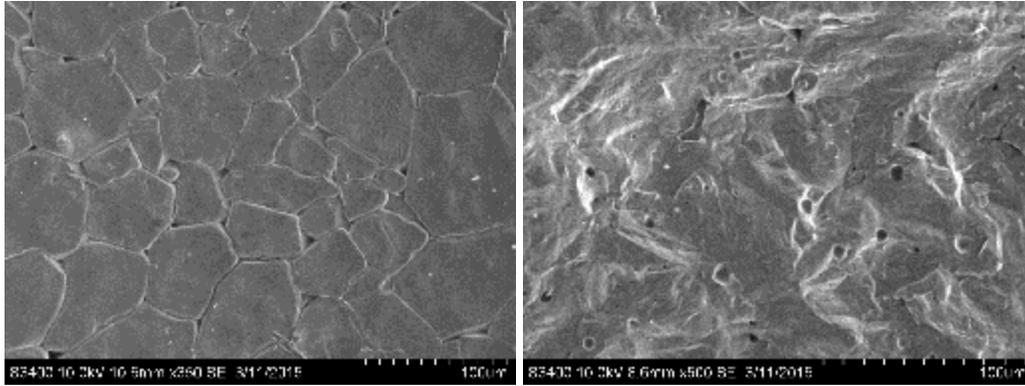
**Fig. 7. SEM microstructure of BT_1 pellet surface (a) and the fracture surface (b)**

In the literature, one can not find the four parameters of barium titanate sinters influencing on the usability properties of multilayer actuators, namely the coefficient of linear expansion, the bending strength, the hardness and the fracture toughness.

The coefficient of thermal expansion for BT_1 material in the range of temperature from 20 to 600°C was equal to $4.75 \cdot 10^{-6}$ 1/K, which should ensure the silver electrode deposition and sintering at about 600°C (Zheng et al., 2012) without any mismatching of thermal behaviours of both barium titanate and silver paint.

The average value of bending strength was determined as 78.1 MPa, which is a relatively high value taking into account that the surface of actuator will not take a great dimension. The Vickers hardness of BT_1 beam was equal to 3.09 GPa, while the fracture toughness has the value 0.83 MPa·m$^{1/2}$, which also possess reasonable values for application the barium titanate sinters as the main part of stacked-disk multilayer actuator.

## CONCLUSIONS

During synthesis of barium titanate powder by solid state method, the monophase material was obtained after the second mechanical activation and calcination. The spherical grains of powder of the size about 500 nm were indicated on SEM microstructure of powder but these small grains have an easy tendency to create agglomerates. The small values of the specific surface area of both powder and granulate as well as the neglecting pores volumes ensured the good properties of material mouldability and finally allowed to receive sinters of high density. Obtainment of the high density of sinters was confirmed by results of the apparent density and additionally photos of microstructure. The purity of powder, granulate and sinters was controlled by determining the phase compositions, which in all cases revealed the almost perfect barium titanate phase.

The four parameters of barium titanate sinter affecting on the actuators usability properties, according to the best of authors knowledge, not found before in the literature, have reasonable values.

Although, the obtained in this approach the sinter microstructure, is very similar to this presented in literature by other authors, in order to receive better result of the barium titanate dielectric constant, the microstructure of sinter must be improved in the direction of the smaller grains getting. This problem will be solved by the thermal treatment of green pellets according to the special designed sintering curves.

**Acknowledgments**
The research leading to these results has received funding from the People Programme (Marie Curie Actions) of the European Union's Seventh Framework Programme FP7/2007-2013/ under REA grant agreement n° PITN-GA-2013- 606878.